\documentclass[reprint,
superscriptaddress,
 amsmath,amssymb,
 aps,
floatfix,
longbibliography
]{revtex4-1}
\usepackage{amsmath}
\usepackage{graphicx}% Include figure files
\usepackage{multirow}
\usepackage{dcolumn}% Align table columns on decimal point
\usepackage{bm}% bold math
\usepackage{xcolor}      

\begin{document}
%setlength{\headheight}{25pt}

%\pagestyle{fancy}

%\rhead{\includegraphics[width=2.5cm]{vch-logo1.png}}

\title{Nonlinear dynamics and magneto-elasticity of nanodrums near the phase transition}

\author{Makars \v{S}i\v{s}kins}
\email{e-mail: makars@nus.edu.sg; f.alijani@tudelft.nl}
\affiliation{
	Department of Precision and Microsystems Engineering, Delft University of Technology,
	\\Mekelweg 2, 2628 CD, Delft, The Netherlands}
\affiliation{Institute for Functional Intelligent Materials, National University of Singapore, \\4 Science
Drive 2, Singapore 117544, Singapore}
\author{Ata Ke\c{s}kekler}
\affiliation{
	Department of Precision and Microsystems Engineering, Delft University of Technology,
	\\Mekelweg 2, 2628 CD, Delft, The Netherlands}

\author{Maurits Houmes}

\affiliation{%
 Kavli Institute of Nanoscience, Delft University of Technology, \\Lorentzweg 1,
 2628 CJ, Delft, The Netherlands
}

\author{Samuel Ma\~{n}as-Valero}
\affiliation{%
 Kavli Institute of Nanoscience, Delft University of Technology, \\Lorentzweg 1,
 2628 CJ, Delft, The Netherlands
}
\affiliation{%
Instituto de Ciencia Molecular (ICMol), Universitat de Val\`{e}ncia, \\c/Catedr\'{a}tico Jos\'{e} Beltr\'{a}n 2, 46980 Paterna, Spain
}%

\author{Eugenio Coronado}
\affiliation{%
Instituto de Ciencia Molecular (ICMol), Universitat de Val\`{e}ncia, \\c/Catedr\'{a}tico Jos\'{e} Beltr\'{a}n 2, 46980 Paterna, Spain
}%

\author{Yaroslav M. Blanter}%
\author{Herre S. J. van der Zant}
\affiliation{%
 Kavli Institute of Nanoscience, Delft University of Technology, \\Lorentzweg 1,
 2628 CJ, Delft, The Netherlands
}

\author{Peter G. Steeneken}
\affiliation{
	Department of Precision and Microsystems Engineering, Delft University of Technology,
	\\Mekelweg 2, 2628 CD, Delft, The Netherlands}
\affiliation{%
	Kavli Institute of Nanoscience, Delft University of Technology, \\Lorentzweg 1,
	2628 CJ, Delft, The Netherlands
}%
\author{Farbod Alijani}
\email{e-mail: makars@nus.edu.sg; f.alijani@tudelft.nl}
\affiliation{
	Department of Precision and Microsystems Engineering, Delft University of Technology,
	\\Mekelweg 2, 2628 CD, Delft, The Netherlands}

\begin{abstract}
Nanomechanical resonances of two-dimensional (2D) materials are sensitive probes for condensed-matter physics, offering new insights into magnetic and electronic phase transitions. Despite extensive research, the influence of the spin dynamics near a second-order phase transition on the nonlinear dynamics of 2D membranes has remained largely unexplored. Here, we investigate nonlinear magneto-mechanical coupling to antiferromagnetic order in suspended FePS$_3$-based heterostructure membranes. By monitoring the motion of these membranes as a function of temperature, we observe characteristic features in both nonlinear stiffness and damping close to the N\'{e}el temperature $T_{\rm{N}}$. We account for these experimental observations with an analytical magnetostriction model in which these nonlinearities emerge from a coupling between mechanical and magnetic oscillations, demonstrating that magneto-elasticity can lead to nonlinear damping. Our findings thus provide insights into the thermodynamics and magneto-mechanical energy dissipation mechanisms in nanomechanical resonators due to the material's phase change and magnetic order relaxation.
\end{abstract}

%Two-dimensional (2D) material membranes are highly advantageous for use as nanodrum resonators due to their rich linear and nonlinear dynamics along with a diversity of available coupling degrees of freedom. Despite extensive research, second-order phase transitions' effect on nonlinear dynamics in 2D membranes remained elusive due to the lack of experimental evidence of corresponding coupling mechanisms

%% Start the actual chapter on a new page.
\maketitle

The mechanical properties of two-dimensional (2D) materials have been extensively studied \cite{Androulidakis2018,Jiang2019} due to their potential for use in a variety of applications, such as sensing \cite{SensorsLemme2020,Jiang2019,Bacteria2022} and energy transduction \cite{Steeneken2021,Guttinger2017,Bachtold2022}. Owing to its superior sensitivity to applied forces, the motion of these membranes can easily be coupled to various degrees of freedom \cite{Steeneken2021,Bachtold2022}, ranging from coupling to photons \cite{Kirchhof2022, EvaWeig2023}, phonons \cite{Mark2016, DeAlba2016,GrapheneDistantLuo2018} and electrons \cite{Chen2015, Sengupta2010,CDWLee2021}, to an interaction between multiple resonators at a distance \cite{GrapheneDistantLuo2018, SiskinsSokolovskaya2021}. Their small mass and ultra-thin nature also makes them highly susceptible to geometric nonlinearities \cite{Davidovikj2017nonlin}, leading to internal resonances \cite{Keskekler2021,Keskekler2022} and various nonlinear dissipation mechanisms  \cite{Mark2016, Keskekler2021, Guttinger2017,Eichler2011} that can dictate their motion dynamics at relatively small amplitudes. 

Recently, there has been a growing interest in using nanomechanical vibrations of 2D materials as practical nodes for inferring elastic and thermodynamic properties of 2D membranes \cite{Steeneken2021}. Examples include nonlinear dynamic characterization of their elastic properties \cite{Davidovikj2017nonlin}, probing magnetic \cite{Siskins2020,FaiMakJiang2020,CGTSiskins2021,Zhang2022,LpezCabrelles2021,Li2022, Maurits2023} and electronic phase transitions \cite{Siskins2020, CDWLee2021}. Among them, the ability of these membranes to detect magnetic phase change in the absence of an applied magnetic field \cite{Siskins2020, CGTSiskins2021,LpezCabrelles2021} has opened up new avenues for developing self-sensitive magnetic nano-electromechanical (NEMS) devices \cite{Steeneken2021,Bachtold2022}. This approach relies on the coupling between the magnetic and mechanical properties of the 2D material, which allows for highly sensitive detection of magnetisation \cite{FaiMakJiang2020,Maurits2023} and thermodynamics of magnetic phases \cite{Siskins2020, CGTSiskins2021}. Furthermore, since these freestanding 2D materials are easily driven to the nonlinear regime of mechanical motion \cite{Davidovikj2017nonlin, Keskekler2021}, the comprehensive studies and analysis of nonlinear dynamics become important given that their magneto-elastic interactions and microscopic dissipation pathways are inherently intricate. 

Here, we explore the effect of magneto-elastic coupling and magnetic order on the nonlinear dynamics of antiferromagnetic membranes made of FePS$_3$-based heterostructures. We study the changes in both nonlinear stiffness and nonlinear damping as a result of the antiferromagnetic phase transition near the N\'{e}el temperature $T_{\rm N}$ of FePS$_3$ \cite{Lee2016,Siskins2020}. Consequently, we describe these experimental observations with a magnetostriction model, revealing and providing a description of the magneto-mechanical dissipation mechanism as a previously unexplored source of nonlinear damping in 2D material membranes.

\begin{figure*}
\begin{center}
	\includegraphics[width=\linewidth]{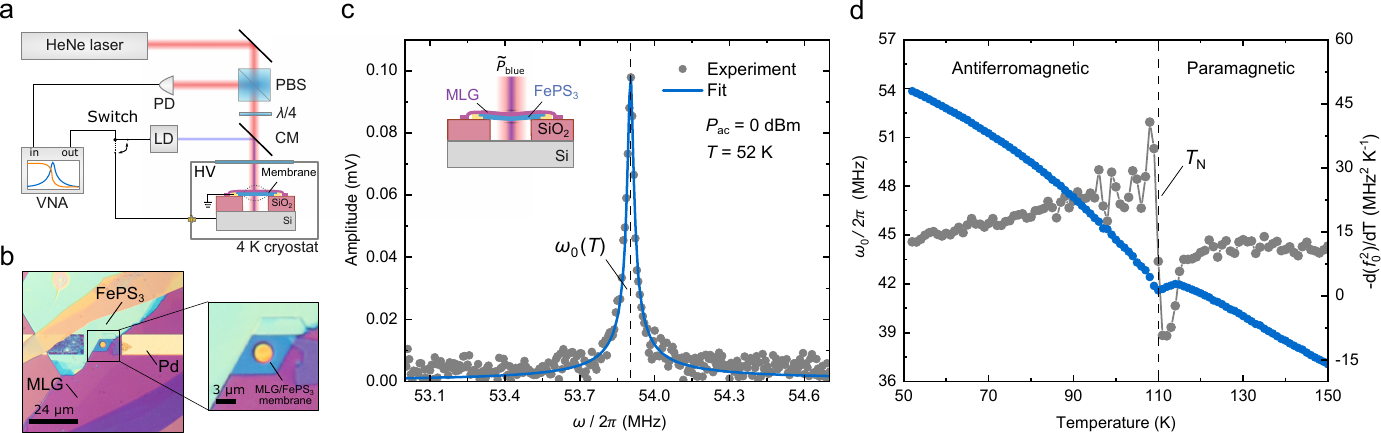}
	\caption{Membrane resonator made of MLG/FePS$_3$ heterostructure. \textbf{a} Schematic of the laser interferometer measurement setup (see Methods). PD is the photodiode, LD - the laser diode, CM - the cold (dichroic) mirror, PBS - the polarized beam splitter, VNA - the vector network analyzer. \textbf{b} Optical image of the sample. \textbf{c} The measured fundamental resonance peak of the membrane (filled grey dots) at $0$ dBm opto-thermal drive. The solid blue line is fit of the linear damped harmonic oscillator model. The inset shows the schematic of the device cross-section. A vertical dashed line indicates extracted $\omega_0$. \textbf{d} The resonance frequency $\omega_0$ as a function of temperature, extracted from the fit similar to (c) (filled blue dots). Connected grey dots are the corresponding derivative of the $f_0^2$. A vertical dashed line indicates $T_{\rm N}$.}
	\label{Fig1}
\end{center}
\end{figure*}

In creating a freestanding membrane, we suspend a $9.5\pm0.6$ nm thin layer of FePS$_3$ over a pre-defined circular cavity with a radius $r=1.5$ $\mu$m in a Si/SiO$_2$ substrate (Fig.~\ref{Fig1}). To improve the thermal conductivity of the FePS$_3$-based heterostructure \cite{Kargar2020} and electrically contact it, we cover the membrane with multi-layer graphene (MLG) of $2.0\pm0.7$ nm thickness which provides an excellent thermal sink \cite{grapheneThermalXu2014,TauDolleman2017}. These MLG/FePS$_3$ heterostructure membranes are then placed in an optical closed-cycle cryostat chamber and cooled to cryogenic temperatures. At a specific temperature $T$ set by the local sample heater, we interferometrically measure the amplitude of the membrane's fundamental mode of vibration $x$ in response to the low-power opto-thermal drive \cite{Siskins2020, Davidovikj2016visual} (see Methods and Fig.~\ref{Fig1}a-c). We then fit the measured resonance peak (grey-filled dots) to the linear harmonic oscillator model (solid blue line) and extract the corresponding resonance frequency $\omega_0(T)=2\pi f_0(T)$, as shown in Fig.~\ref{Fig1}c.

Following this procedure, we measure $\omega_0(T)$ in the temperature range from $52$ to $150$ K as shown in Fig.~\ref{Fig1}d. In the vicinity of $T\sim 110$ K (vertical dashed line in Fig.~\ref{Fig1}d) the resonance frequency $\omega_0(T)$ exhibits the antiferromagnetic-to-paramagnetic phase transition-related anomaly. This becomes even more prominent in the temperature derivative of $f_0^2(T)$ (filled grey dots in Fig.~\ref{Fig1}d) - a quantity which is related to specific heat $c_v(T)$ of the material through thermal expansion coefficient and Grüneisen parameter \cite{Siskins2020}. Thus, the temperature of the discontinuity in $-\frac{\text{d}f_0^2(T)}{\text{d}T}$ can be used as a measure of $T_{\rm N}$ at the transition from ordered to disordered magnetic state \cite{Siskins2020, CGTSiskins2021}. This is further supported by the fact that the measured $T_{\rm N}$ also corresponds to a peak in inverse quality factor $Q^{-1}(T)$ (see Supplementary Note 1), which is expected to arise near the phase transition temperature \cite{Siskins2020, CGTSiskins2021, SiskinsSokolovskaya2021}.

\begin{figure*}[ht]
\begin{center}
	\includegraphics[width=0.95\linewidth]{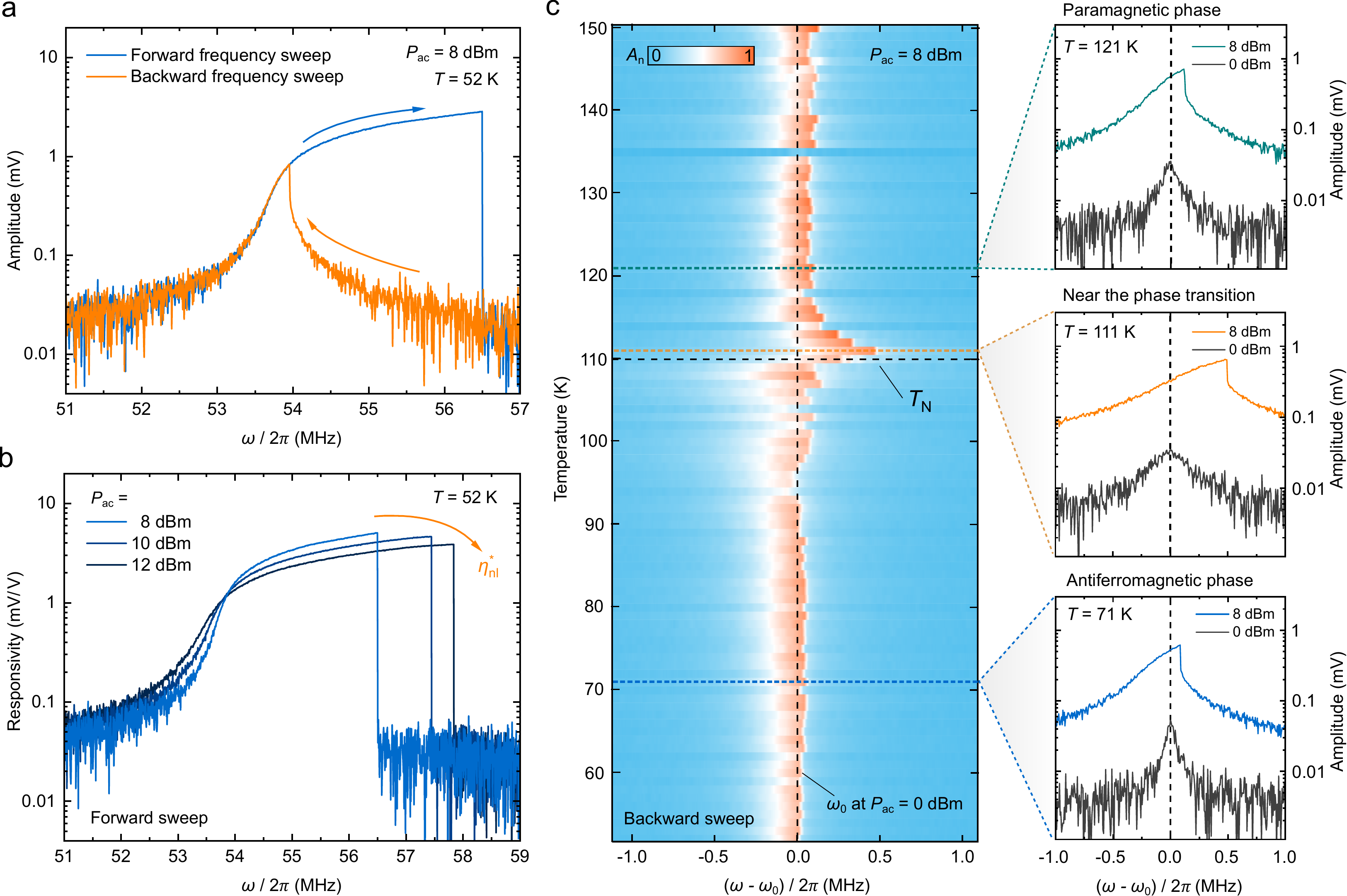}
	\caption{Nonlinear dynamics of MLG/FePS$_3$ membrane. \textbf{a} The measured Duffing response and amplitude branches of the resonance peak from Fig.~\ref{Fig1}c at higher excitation power ($P_{\rm ac}=8$ dBm). \textbf{b} The measured resonance peak responsivity, i.e. drive power-normalized amplitude,  at $8$, $10$ and $12$ dBm for the same temperature from (a) indicating the presence of nonlinear damping $\eta^*_{\rm nl}$.\textbf{c} Left panel: Colour map of the normalized amplitude measured as a function of temperature for backward frequency sweeps with respect to the linear resonance frequency $\omega_0(T)$ shown in Fig.~\ref{Fig1}d. The N\'{e}el temperature $T_{\rm N}$ from Fig.~\ref{Fig1}d is indicated with a black dashed horizontal line. Right panel: the measured frequency response around $\omega_0$ corresponding to dashed line cuts from the left panel for $P_{\rm ac}=0$ and $8$ dBm at three temperature points corresponding to different magnetic phases. }
	\label{Fig2}
\end{center}
\end{figure*}

After characterising the dynamics of the membrane in the linear regime and at a low opto-thermal driving force, we increase the drive from $0$ to $8$ dBm to achieve higher force levels and observe features of the nonlinear motion \cite{Davidovikj2017nonlin}. Fig.~\ref{Fig2}a displays an apparent Duffing effect measured at $T=52$ K and $8$ dBm, revealing bi-stable amplitude behaviour that depends on the direction of the frequency sweep. By further increasing $P_{\rm ac}$, we observe a corresponding decrease in responsivity of the resonance peak, shown in Fig.~\ref{Fig2}b. This indicates the presence of nonlinear damping in the system, which becomes apparent at high amplitudes of motion \cite{Keskekler2021}. We measure the amplitude of membrane motion around $\omega_0(T)$ at $8$ dBm of drive in the temperature range from $52$ to $150$ K and plot it in Fig.~\ref{Fig2}c with respect to measured $\omega_0(T)$ in the linear regime from Fig.~\ref{Fig1}d. Two noteworthy observations can be made: first, the position of the resonance peak at a higher driving power is shifted to higher frequencies near $T_{\rm N}$, indicating a change in linear membrane stiffness $k_1$, corresponding to a change in the strain \cite{Siskins2020}; secondly, the peak amplitude of the Duffing response and its associated frequency changes depending on the magnetic state of the membrane with the largest effect near $T_{\rm N}$, indicating a change in nonlinear membrane stiffness $k_3$ \cite{Davidovikj2017nonlin} (see Fig.~\ref{Fig2}c and Supplementary Note 2). We have also performed control experiments on multiple samples using both optical and electrical excitation, where an AC voltage $V_{\rm ac}$ signal is applied between the Si backgate of the chip and the conducting top layer of MLG.
Since we obtain similar results for the electrostatic drive as for optothermal drive we conclude that the reported observations are intrinsic to the resonator and not related to the driving mechanism (see Supplementary Note 3).

\begin{figure*}[hp]
\begin{center}
	\includegraphics[width=0.84\linewidth]{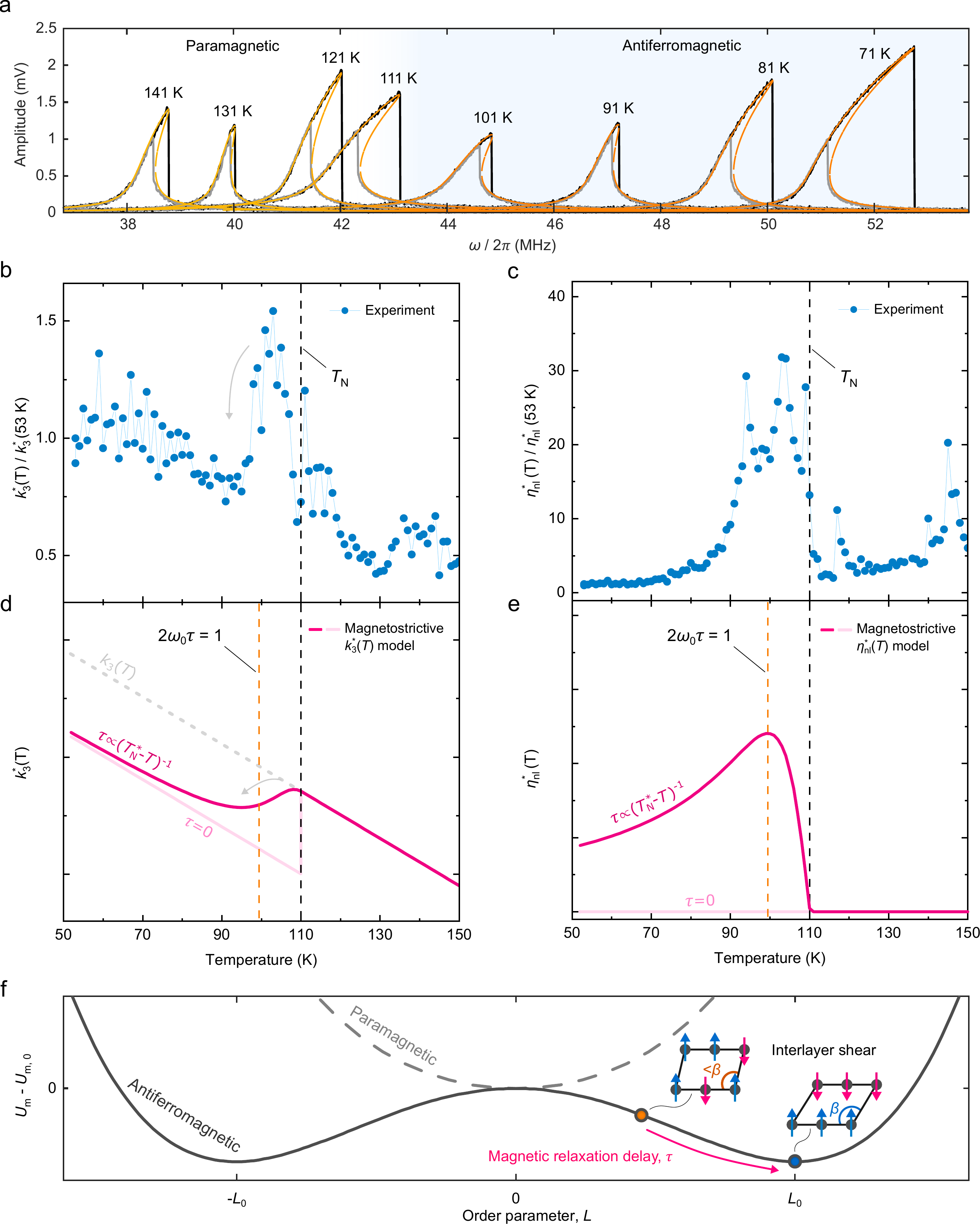}
	\caption{Temperature dependence of the nonlinear stiffness $k^*_3$ and nonlinear damping $\eta^*_{\rm nl}$ of a magnetostrictive membrane resonator. \textbf{a} Fit of the equation~(\ref{eq:amplitude}) (orange solid lines) to the measured amplitude for forward frequency sweep (black solid lines) and backward frequency sweeps (grey solid lines) at $P_{\rm ac}=10$ dBm and temperature point indicated. The light blue region schematically indicates the antiferromagnetic phase of the membrane. \textbf{b} Connected blue dots - the measured $k^*_3(T)$, extracted from the fit similar to (a), normalized by the value at $53$ K. \textbf{c} Connected blue dots - the measured $\eta^*_{\rm nl}(T)$, extracted from the fit similar to (a), normalized by the value at $53$ K. \textbf{d} and \textbf{e} Solid magenta lines - the nonlinear stiffness and nonlinear damping model of equations~(\ref{k3star}) and (\ref{nldstar}) respectively at $\omega=\omega_0(T)$ from Fig.~\ref{Fig1}d and $\tau^{-1}(T)=2\kappa a\left(T^*_{\rm N}-T\right)$ from Zhou et al \cite{Zhou2022} for $h=9.5$ nm, solid light magenta lines - the model of equations~(\ref{k3star}) and (\ref{nldstar}) at the same $\omega=\omega_0(T)$ and $\tau=0$ (see Supplementary Note 5). The dashed light grey line in (d) - non-magnetic $k_3(T)$ slope extracted by a linear fit to $T>110$ K region in (b). Vertical dashed orange lines in (d) and (e) - the temperature point at which $2\omega_0\tau=1$, producing a maximum in the nonlinear damping $\eta^*_{\rm nl}(T)$. \textbf{f} Schematic of the magnetic free energy of the system with un-relaxed (orange dot) and relaxed ground (blue dot) states indicated. Magnetic sub-lattice relaxation dynamics is accompanied by a slow interlayer shear deformation with a change in the monoclinic angle $\beta$ \cite{Zhou2022, Zong2023} schematically indicated in insets, which hypothetically may have the dominant contribution to $\tau$.}
	\label{Fig4}
\end{center}
\end{figure*}

To qualitatively interpret the experimental findings as a function of temperature, we utilize a dedicated algorithm to fit the measured nonlinear response at different temperatures in the vicinity of $T_{\rm{N}}$. Our approach involves fitting the experimental data with the Duffing-van der Pol equation (see equation~(\ref{eq:amplitude}) in Methods and Supplementary Note 4), as depicted in Fig.~\ref{Fig4}a. To avoid an over-parameterised fitting procedure and reduce the uncertainty of the fit, we first extract quality factors $Q(T)$ and $\omega_0(T)$ from the linear resonance peak at low drive levels. Next, we extract the relative driving force $F_{\omega}(T)$ by fitting the off-resonance response to a harmonic oscillator model. After obtaining all the linear parameters, we obtain the Duffing term $k^*_3(T)$ at $P_{\rm ac}=10$ dBm (Fig.~\ref{Fig4}b) from the slope of the backward frequency sweep response, which is unaffected by nonlinear damping \cite{Keskekler2021}. Consequently, we fix this value to fit the forward frequency sweep response, thus extracting the van der Pol-type nonlinear damping term $\eta^*_{\rm nl}(T)$ using an optimizer algorithm (see Supplementary Note 4). We plot the extracted nonlinear damping term in Fig.~\ref{Fig4}c for the temperature range $52-150$ K. As seen from the results of the fit to experimental data, at a higher driving power and as the temperature decreases, a sharp drop is observed in $k^*_3(T)$ at $T<T_{\rm N}$. This feature is also accompanied by a peak in $\eta^*_{\rm nl}(T)$ at approximately the same temperature. 

Pronounced features in both $k^*_3(T)$ and $\eta^*_{\rm nl}(T)$ close to $T_{\rm N}$ shown in Fig.~\ref{Fig4}b and c indicate the softening of nonlinear stiffness as well as a prominent increase in the nonlinear dissipation in the antiferromagnetic phase of FePS$_3$, suggesting the magnetic origin of the effect. Therefore, to underpin the influence of magneto-mechanical coupling on our observations, we model the system by considering the elastic potential energy as a function of the membrane displacement at its centre $U_{\rm el}$ and the magnetic free energy $U_{\rm m}$ of FePS$_3$, coupled via spontaneous magnetostriction $U_{\rm ms}$ \cite{Siskins2020, Maurits2023, Landau1984} (see Supplementary Note 5):
\begin{equation}\label{eq:totalenergy}
\begin{split}
    U_{\rm T}=&U_{\rm el}+U_{\rm m}+U_{\rm ms}\\
    =&\left[\frac{k_1}{2}x^2+\frac{k_3}{4}x^4\right]+\\&\left[U_{\rm{m},0}+\frac{a\left(T-T_{\rm{N}}\right)}{2}L^2+\frac{B}{4}L^4\right]+\left[\frac{\lambda_{ij}\sigma_{ij}(x)}{2} L^2\right],
\end{split}
\end{equation}
where $\sigma_{ij}(x)$ is the amplitude-dependent stress tensor, $L$ the antiferromagnetic order parameter in the direction of the easy-axis of FePS$_3$, $\lambda_{ij}$ the magnetostriction tensor, $U_{\rm{m},0}$ is the magnetic energy in the paramagnetic state, and $a$, $B$ are phenomenological positive constants \cite{Landau:1937obd, Landau1984}. By minimizing equation~(\ref{eq:totalenergy}) with respect to $L$ at a static deformation $\omega=0$, the ground state order parameter $L_0$ is obtained (see Methods and Supplementary Note 5). When the membrane is in motion and the magnetic system is out of equilibrium, the order parameter is stress- and time-dependent as $L(t)\simeq L_0 + L_{\omega}(t)$. Subsequently, the rate at which $L(t)$ approaches the ground state $L_0$ (Fig.~\ref{Fig4}f) is described by the kinetic equation \cite{LandauTAU1954,Belov1960,BelovFerromagnet1959}:
\begin{equation}\label{eq:kineticeq}
    \frac{\text{d}L}{\text{d}t}=-\kappa\frac{\partial U_{\rm T}}{\partial L},
\end{equation}
where $t$ is the time and $\kappa$ the phenomenological kinetic coefficient, which we assume to be temperature-independent for simplicity. 

We further describe the driven coupled magneto-mechanical system by linearizing equation~(\ref{eq:kineticeq}) near $L_0$ together with obtaining the equation of motion associated with the generalized coordinate $x$. In doing that, we define the Lagrangian $\mathcal{L}=\frac{1}{2}m\dot{x}^2 - U_{\rm{T}}$ and use the Euler-Lagrange equations to obtain the system of coupled dynamic equations: 
\begin{align}
\label{eq:lomega} 
    \dot L_{\omega}+\frac{L_{\omega}}{{\tau}}+ \lambda \kappa L_0 \sigma_{\omega}=0,&\\\label{eq:lomega2} 
    m\ddot{x}+k_1x+k_3x^3+\frac{\lambda}{2}L^2&\frac{\partial \sigma(x)}{\partial x}=F_{\omega}\cos{\left(\omega t\right)}\\\nonumber&-\left(\frac{m\omega_0}{Q}+\eta_{\rm nl}x^2\right)\dot{x},
\end{align}
where $\sigma=\sigma_0+\sigma_{\omega}$ with static $\sigma_0$ and dynamic $\sigma_{\omega}$ stress contributions, $F_{\omega}$ the amplitude of periodic driving force and $\tau=\left[2\kappa a\left(T^*_{\rm N}-T\right)\right]^{-1}$ the magnetic relaxation time constant of FePS$_3$ layer \cite{LandauTAU1954, Belov1960, Zhou2022} (see Methods and Supplementary Note 5). Typically fast magnetic relaxations in antiferromagnets are of the order of picoseconds  \cite{Nemec2018, Afanasiev2020, CoPS3Tau2023}. However, in the case of FePS$_3$ long nanosecond-scale relaxation times are required to relax the magnetic sub-lattice near $T_{\rm{N}}$ due to the strongly coupled ordering of spins to the slow process of interlayer shear (Fig.~\ref{Fig4}f) \cite{Zhou2022,Zong2023}.  We hypothesise that the slow spin-shear relaxation mechanism in FePS$_3$ \cite{Zhou2022, Zong2023} may have the dominant contribution to the magnetic time constant $\tau$ of equation~(\ref{eq:lomega}), and hereinafter consider the experimentally measured spin-shear $\tau(T)$ from the work of Zhou et al \cite{Zhou2022} (see Supplementary Note 5). The $L_{\omega}$ term then induces oscillations in $L$, which can lag the membrane motion at sufficiently large $\tau$ \cite{Zhou2022,Zong2023} producing a delay in the coupled magneto-mechanical system. 

Solving the coupled system of equations~(\ref{eq:lomega}) and (\ref{eq:lomega2}) using the harmonic balance method, we obtain the steady-state amplitude-frequency response (see Methods and Supplementary Note 5). As a direct consequence, when the membrane is in motion, the linear and nonlinear stiffness as well as nonlinear damping coefficients are renormalized by additional magnetic terms, which yield the following steady-state equation of Duffing-van der Pol type \cite{Keskekler2021}:
\begin{equation} \label{eq:motion}
    %m\ddot{x}+k^*_1x+k^*_3x^3=F_{\omega}\cos{\left(\omega t\right)}-\frac{m\omega_0}{Q}\dot{x}-\eta^*_{\rm nl}x^2\dot{x},
    \begin{aligned}
	\left(\frac{3k^*_3}{4}a_{\rm s}^3 +m(\omega_0^2-\omega^2)a_{\rm s}\right)^2+\left(\eta^*_{\rm{nl}}a_{\rm s}^3+\frac{m\omega_0}{Q}a_{\rm s} \right)^2\omega^2=F_\omega^2,
\end{aligned}
\end{equation}
in which $a_{\rm s}$ is the steady-state amplitude, $m$ the effective mass of the resonator, $F_{\omega}$ the drive force amplitude, $m\omega^2_0=k^*_1=k_1+\lambda L_0^2\frac{Ec_3}{2r^2}$ the renormalized linear stiffness, $k^*_3$ the renormalized nonlinear stiffness:
\begin{equation}\label{k3star}
k^*_3=
\begin{cases}
k_3-\frac{\lambda^2}{12B}\frac{E^2c_3^2}{r^4}\frac{1}{1+4\omega^2\tau^2} & T<T_{\rm{N}}^{*} \\
k_3 & T>T_{\rm{N}}^{*},
\end{cases}
\end{equation}
and $\eta^*_{\rm nl}$ the magnetic nonlinear damping term of van der Pol type \cite{Schmid2016,Keskekler2021}:
\begin{equation}\label{nldstar}
\eta^*_{\rm nl}=
\begin{cases}
\eta_{\rm nl}+\frac{\lambda^2}{2B}\frac{E^2c_3^2}{r^4}\frac{\tau}{1+4\omega^2\tau^2} & T<T_{\rm{N}}^{*} \\
\eta_{\rm nl} & T>T_{\rm{N}}^{*},
\end{cases}
\end{equation}
where $k_3$ is the non-magnetic nonlinear stiffness, $\eta_{\rm nl}$ is the non-magnetic nonlinear damping, $E$ the Young's modulus and $c_3$ the geometric numerical factor that also depends on membrane’s Poisson ratio \cite{Davidovikj2017nonlin}.

%\begin{figure}
%\begin{center}
%	\includegraphics[width=0.80\linewidth]{Fig3_v4.pdf}
%	\caption{Schematic of the relaxation mechanism of antiferromagnetic order parameter $L$. \textbf{a} Schematic of the magnetic free energy of the system with un-relaxed (orange dot) and relaxed ground (blue dot) states indicated. \textbf{b} Schematic of the resulting dynamic shift in $T^*_{\rm N}$ and corresponding modulation of $L$ as a function of $x$ and $\omega$.}
	\label{Fig3}
%\end{center}
%\end{figure}

Renormalization of $k^*_1$ and $k^*_3$ leads to two important consequences. First, since $L$ shall turn to zero above the phase transition temperature, strain reduces the transition temperature as $T_{\rm N}^*=T_{\rm N}-\frac{\lambda_{ij} \sigma_{ij}(x)}{a}$, which was previously demonstrated by applying a static external force \cite{Siskins2020}. Likewise, at high amplitude oscillations, the dynamic change of the stress via a modulated force results in an additional effective static strain and related stress term (see Supplementary Note 5), which can reach up to $0.03\%$ in similar systems \cite{DynamicStrainZhang2020} and accordingly reduce $T_{\rm N}^*$ of FePS$_3$ by a few Kelvins \cite{Siskins2020}. This produces a corresponding change in $k_1$ and a shift of the phase transition-related feature in $\omega_0(T)$ near $T_{\rm N}$, consequently causing the above-mentioned shift of the resonance curve with respect to $\omega_0$ at a higher driving power in Fig.~\ref{Fig2}c (see Supplementary Note 5). The contribution of the order parameter on effective linear stiffness $k^*_1\propto L_0^2$ is studied and described in details in a previous work \cite{Maurits2023}.

Second, unlike the renormalization of $k_1$, which is independent of dynamics of the order parameter $L$, consequences for nonlinear parameters $k_3$ and $\eta_{\rm nl}$ arise from the modulation of the order parameter. As a result, both $k^*_3$ and $\eta^*_{\rm nl}$ are functions of a characteristic delay of the coupled dynamic system described by $\tau$ and $\omega$. As follows from equation~(\ref{k3star}), $k^*_3$ starts to decrease with $\delta k^*_3 \propto -\frac{1}{1+4\omega^2\tau^2}$ when $T<T^*_{\rm N}$. The same magnetic contribution also leads to substantial nonlinear damping $\eta^*_{\rm nl}$ at $T<T^*_{\rm N}$, which scales as $\delta \eta^*_{\rm nl}\propto \frac{\tau}{1+4\omega^2\tau^2}$ and peaks at $2\omega\tau\simeq1$ (see Methods and Supplementary Note 5). This behaviour can be understood intuitively: magnetostriction mediates the exchange of the membrane's mechanical energy with a coupled magnetic reservoir, which can happen twice for one period of motion due to symmetrical modulation of stress in the up-down geometry of its deflection. If membrane oscillations are much faster than the energy exchange rate to a coupled magnetic reservoir, i.e $2\omega \gg 1/\tau$, there is not enough time for it to relax and dissipate energy. On the contrary, when the oscillations are at a much slower timescale $2\omega \ll 1/\tau$, the energy exchange follows the oscillations with a negligible delay, again resulting in minimal dissipation \cite{Schmid2016}. Thus, the nonlinear damping due to coupling to the order parameter peaks when the relaxation delay is significant and $2\omega\tau\simeq1$.

Subsequently, we plot the derived magnetostrictive model of equations~(\ref{k3star}) and (\ref{nldstar}) for $\omega=\omega_0$ in Fig.~\ref{Fig4}d and e, next to the measured $k^*_3(T)$ and $\eta^*_{\rm nl}(T)$ in Fig.~\ref{Fig4}b and c. We assume the non-magnetic Duffing constant $k_3$ to be temperature dependent, providing the additional background-slope in $k^*_3(T)$ below and above $T^*_{\rm N}$. As shown in Fig.~\ref{Fig4}b with a solid magenta line, equation~(\ref{k3star}) reproduces the measured decrease of $k^*_3$ in the proximity of $T^*_{\rm N}$. At the same time, the same model in Fig.~\ref{Fig4}c reproduces the measured peak in $\eta^*_{\rm nl}(T)$ at $2\omega_0\tau=1$. Notably, in a hypothetical case where $\tau$ is sufficiently small, i.e. $\tau=0$ in equations~(\ref{k3star}) and (\ref{nldstar}), the model predicts the discontinuous decrease in $k^*_3(T)$ at $T^*_{\rm N}$, while the magnetic contribution to $\eta^*_{\rm nl}(T)$ completely vanishes as shown in Fig.~\ref{Fig4}d-e with light magenta lines. 

In discussing the physical interpretation of the origin of this nonlinear damping, its microscopic mechanism should be envisioned as a consequence of a nonlinear oscillator's excited vibrational modes scattering off its own magnetic energy reservoir \cite{Mark1975, Mark2016}. This interaction then is accompanied by the energy transfer of two oscillation quanta ($2\omega_0$) for nonlinear damping \cite{Mark1975}. Importantly, a rather general form of free energy equation and low order of the coupling term suggests that similar effects may appear in systems with other types of non-magnetic phase transitions, for instance, charge density wave \cite{Siskins2020} or coupling the mechanical motion to an electronic energy reservoir. Interestingly, this mechanism also finds its macroscopic similarities to magnetic internal friction arising due to a delay in Young's modulus relaxation near $T_{\rm N}$ which occurs in large-scale bulk of magnetic solids \cite{MaterialsBookSpringer, BelovFerromagnet1959, Belov1960, EAnomalyHausch1973, EAnomaly2Hausch1973, EAnomalyPostolache2000}. However, the crucial distinction at the nanoscale is that it affects different mechanical properties at twice the resonance frequency. Our analysis predicts the observed nonlinear effect in this system appearing solely as a result of modulation of the antiferromagnetic order parameter $L$ with dynamic strain via magnetostriction, delayed by a suggested spin-shear relaxation $\tau$ \cite{Zhou2022,Zong2023} (see Fig.~\ref{Fig4}f). This is supported by a case of $\lambda=0$ eliminating all magnetic contributions to both $k^*_3$ and $\eta^*_{\rm nl}$. Perhaps, some additional effects may also contribute to a part of our observation. One such noteworthy effect is a similar relaxation due to thermoelasticity \cite{Siskins2020}. Yet, the latest experiments show that thermal relaxation time-scales in membranes of FePS$_3$ are up to two orders of magnitude slower \cite{Gabriele2023} than spin-shear relaxation-related $\tau$ considered in this work for comparable sample thicknesses \cite{Zhou2022}. Therefore, the presence of substantial linear thermoelastic damping and the probed nonlinear damping near $T_{\rm{N}}$ are not a direct consequence of each other \cite{Siskins2020,Gabriele2023}. This is further justified by the fact that magneto-mechanical coupling and the associated relaxation mechanism does not lead to any linear damping terms analytically (See Supplementary Note 5). Another contribution may come from nonlinear effects, like nonlinearities in optothermal response \cite{Barton2012,AmpDolleman2017}, and resulting nonlinear terms in the magnetostrictive actuation force \cite{Gabriele2023} that may affect the change near the magnetic phase transition. Nevertheless, quantitatively confirming either of these hypotheses would require further experimental evidence.

In conclusion, we demonstrated the nonlinear nanomechanical coupling to antiferromagnetic order in FePS$_3$-based heterostructure membranes. We provide both experimental evidence and theoretical descriptions of the mechanism responsible for the renormalization of the nonlinear parameters. We demonstrate a previously unexplored magneto-mechanical dissipation mechanism supported by a microscopic theory that accounts for magnetostriction, which strongly affects the nonlinear dynamics of magnetic membranes, even in the absence of a magnetic field, near the phase transition temperature. We anticipate that our discoveries offer a new understanding of the thermodynamics and energy dissipation mechanisms related to magneto-mechanical interactions in 2D materials, which is important for future studies of more intricate magnetic systems, like 2D quantum phases and moir\'{e} magnets \cite{Burch2018}, as well as the development of novel magnetic NEMS and spintronic devices.

\section*{Methods}
\subsection*{Sample fabrication and characterisation}
We pre-pattern a diced Si/SiO$_2$ wafer with circular holes using e-beam lithography and reactive ion etching. The holes have a radius of $r=1.5$~$\mu$m and a cavity depth of $285$ nm, and the SiO$_2$ layer acts as electrical insulation between the 2D material membranes and the bottom Si electrode. For electrostatic experiments, Pd electrodes are patterned on top of Si/SiO2 chips using a lift-off technique to establish electrical contact with some samples. To create suspended membranes, thin flakes of FePS$_3$ and graphite crystals are mechanically exfoliated and transferred onto the chip using the all-dry viscoelastic stamping method \cite{CastellanosGomez2014Stamping} immediately after exfoliation. Flakes of van der Waals crystals are exfoliated from high-quality synthetically grown crystals with known stoichiometry, and deterministic stacking is performed to form heterostructures. To prevent degradation, samples are kept in an oxygen-free or vacuum environment directly after the fabrication. Atomic Force Microscopy (AFM) height profile scans and inspection are performed in tapping mode on a Bruker Dimension FastScan AFM. We typically use cantilevers with spring constants of $k_{\rm c}=30-40$~N~m$^{-1}$ for inspection. Error bars on reported thickness values are determined by measuring multiple profile scans of the same flake. 

\subsection*{Laser interferometry measurements}
The sample is mounted on a $x-y$ piezo-positioning stage inside a dry optical 4 K cryostat Montana Instruments Cryostation s50. Temperature sweeps are carried out using a local sample heater at a rate of $\sim3$~K~min$^{-1}$ while maintaining the chamber pressure below $10^{-6}$~mbar. During data acquisition, the temperature is maintained constant with $\sim10$~mK stability. A power-modulated blue diode laser with a wavelength of $405$ nm is used to optothermally excite the membrane's motion, and the resulting membrane displacement is measured using an interferometric detection with a He–Ne laser beam of $632$ nm. The interferometer records the interfering reflections from the membrane and the Si electrode underneath, and the data is processed by a vector network analyzer Rohde \& Schwarz ZNB4. All measurements are conducted with incident laser powers of $P_{\rm red}\leq8$~$\mu$W and $P_{\rm blue}\leq 35$~$\mu$W, with a laser spot size of $1$~$\mu$m. To ensure accuracy in the data acquisition, it is verified that resonance frequency changes due to laser heating are insignificant for all membranes for $P_{\rm ac}\leq 15$~dBm. 
 
\subsection*{Derivation of order parameter dynamics}
In the derivation of antiferromagnetic order parameter relaxation dynamics, we follow closely the approach of Landau-Khalatnikov \cite{LandauTAU1954} and Belov-Kataev-Levitin \cite{BelovFerromagnet1959, Belov1960}. For simplicity, we assume the bi-axial in-plane membrane stress $\sigma(x)=\sigma_{xx}=\sigma_{yy}$. First, we derive $\sigma(x)$, assuming $x=a_{\rm{s}}\cos(\omega_0 t)$, as \cite{Davidovikj2017nonlin, DynamicStrainZhang2020}:
\begin{equation} \label{eq:strain_methods}
\begin{split}
\sigma(x)=&\sigma_{\rm{p}}+\frac{Ec_3}{2r^2}x^2\\
=&\left[\sigma_{\rm{p}}+\frac{Ec_3}{4r^2}a_{\rm{s}}^2\right]+\left[\frac{Ec_3}{4r^2}a_{\rm{s}}^2\cos(2\omega_0 t)\right]
\\=&\sigma_{0}+\sigma_{\omega}(t)
\end{split}
\end{equation}
where $a_{\rm{s}}$ is the steady-state amplitude, $\sigma_{\rm{p}}$ the pre-stress in the membrane due to the fabrication process, $\sigma_{0}$ the static and $\sigma_{\omega}(t)$ the dynamic stress terms. Then, we derive $L_0$, antiferromagnetic order parameter ground state, by minimizing the total energy of the magneto-mechanical system~(\ref{eq:totalenergy}) with respect to $L$ at constant bi-axial stress $\sigma_0$ such that:
\begin{equation} 
    \frac{\partial U_{\rm T}}{\partial L}=\frac{\partial \left(U_{\rm m}+U_{\rm ms}\right)}{\partial L}=0,
\end{equation}
resulting in
\begin{equation} 
    L^2_0=\frac{a(T_{\rm N}-T)-\lambda\sigma_0}{B}=\frac{a(T^*_{\rm N}-T)}{B},
\end{equation}
where $\lambda$ is a specific magnetostriction coefficient of $\lambda_{ij}$ tensor that describes coupling of bi-axial in-plane membrane stress $\sigma_0$ to order parameter $L_0$ in the direction of the easy axis.

Using this result we linearize the unrelaxed $L$ as $L\simeq L_0+L_{\omega}$, where $L_{\omega}$ is the time- and amplitude-dependent dynamic term. When the membrane is in motion and $L$ is out of the equilibrium, the rate of relaxation of $L$ to the equilibrium $L_0$ is set by the kinetic equation~(\ref{eq:kineticeq}), which using equation~(\ref{eq:strain_methods}) leads to:
\begin{equation} 
    \frac{dL}{dt}=\frac{dL_{\omega}}{dt}=-\kappa\frac{\partial \left(U_{\rm m}+U_{\rm ms}\right)}{\partial L}.
\end{equation}
This equation can be simplified by Taylor expansion around $L_0$, and assuming $L_{\omega}\ll L_0$, as follows: 
\begin{equation} 
    \frac{dL_{\omega}}{dt}=\dot{L}_{\omega}\simeq-\kappa\left[2BL^2_0L_{\omega}+\lambda L_0\sigma_{\omega}(t)\right],
\end{equation}
which rearranges to equation~(\ref{eq:lomega}), by taking $\tau=\frac{1}{2\kappa a\left(T^*_{\rm N}-T_{\rm N}\right)}$ \cite{LandauTAU1954}.

%For further details of the derivation and fitting procedure see Supplementary Note 5.  

\subsection*{Amplitude of nonlinear resonance peak}
We start by solving the first-order differential equation~(\ref{eq:lomega}) to obtain the steady-state solution for $L_{\omega}$ in terms of $\tau$:
\begin{equation}\label{Ls}
	L_{\omega, \rm{ss}}= -\lambda\kappa L_0  \frac{Ec_3}{4r^2} \frac{\tau  \left[\cos{(2\omega t)}+2 \tau \omega \sin{(2 \omega t)} \right]}{\left(1+4 \tau^2 \omega^2\right)}a_{\rm s}^2.
\end{equation}
We keep the assumption of periodic motion in the form of $x=a_{\rm s}\cos{\omega t}$ and plug in the steady state solution in equation~(\ref{eq:lomega2}) such that $L_{\omega}=L_{\omega, \rm{ss}}$. 

Next we use harmonic balance method to obtain the amplitude-frequency equation~(\ref{eq:motion}), considering only the fundamental harmonic $\omega_0$ (see Supplementary Note 5):
\begin{equation}\label{eq:amplitude}
\begin{split}
a_{\rm{s}}^6 &\left(\frac{9 \gamma^2}{16}+\frac{\xi_{\rm{nl}}^2 \omega^2}{16}\right)+a_{\rm{s}}^4 \left(\frac{3 \gamma(\omega_0^2-\omega^2)}{2}+\frac{\xi_{\rm nl}\omega_0\omega^2}{2Q}\right)+\\
a_{\rm{s}}^2 &\left((\frac{\omega_0\omega}{Q})^2 + (\omega_0^2-\omega^2)^2\right)=\left(\frac{F_{\omega}}{m}\right)^2,
\end{split}
\end{equation}
where $\omega^2_0=\frac{1}{m} \left(k_1+\lambda L_0^2\frac{Ec_3}{2r^2}\right)$ is the re-normalized resonance frequency, $\gamma=\frac{k^*_3}{m}$ the mass-normalized Duffing coefficient and $\xi_{\rm nl}=\frac{\eta^*_{\rm nl}}{m}$ the mass-normalized nonlinear damping coefficient with $k^*_3$ and $\eta^*_{\rm nl}$ from equations~(\ref{k3star}) and (\ref{nldstar}), respectively.

For further details of the derivation and fitting procedure see Supplementary Note 5.

\subsection*{Data availability}
All data supporting the findings of this article and its Supplementary Information will be made available upon request to the authors.

\subsection*{Acknowledgments}
The authors would like to thank Prof. Mark Dykman for support and fruitful discussions about magnetostriction and nonlinear damping. 
M.\v{S}., A.K. and F.A. acknowledge funding from European Union’s Horizon 2020 research and innovation program under Grant Agreement $802093$ (ERC starting grant ENIGMA). M.\v{S}. acknowledges funding from the Ministry of Education, Singapore, under its Research Centre of Excellence award to the Institute for Functional Intelligent Materials, Project No.~EDUNC-$33$-$18$-$279$-v$12$. M.J.A.H., H.S.J.v.d.Z. and P.G.S. acknowledge funding from the European Union's Horizon 2020 research and innovation program under grant agreement number $881603$. Y.M.B and H.S.J.v.d.Z. acknowledge support from Dutch National Science Foundation (NWO). S.M.-V., E.C. acknowledge funding from the European Union (ERC AdG Mol-2D $788222$, ERC StG 2D-SMARTiES $101042680$ and FET OPEN SINFONIA $964396$), the Spanish MCIN (Project 2DHETEROS PID2020-117152RB-100 and Excellence Unit ”Maria de Maeztu” CEX2019-000919-M), and the Generalitat Valenciana (PROMETEO Program and APOST Grant CIAPOS/2021/215 to S.M.-V.).

\subsection*{Author contributions}
M.\v{S}., A.K. and M.J.A.H. performed the laser interferometry measurements and fabricated and inspected the samples. M.J.A.H. fabricated the substrates. S.M.-V. synthesized and characterized the FePS$_{3}$ crystals, supervised by E.C. A.K. developed the fitting algorithm. M.\v{S}., A.K., Y.M.B. and F.A. analysed the experimental data and developed a theoretical model. H.S.J.v.d.Z., P.G.S. and F.A. supervised the project. The paper was jointly written by all authors with a main contribution from M.\v{S}. All authors discussed the results and commented on the paper.

\subsection*{Competing interests}
The authors declare no competing interests.

\bibliographystyle{naturemag}
%\bibliography{tau}% Produces the bibliography via BibTeX.

\end{document}